\documentclass[useAMS,usenatbib]{mn2e}

\usepackage{graphicx}
\usepackage{epsfig}
\usepackage{textcomp}
\usepackage{widetext}
\usepackage{amsmath}
\allowdisplaybreaks[1]
\title
{On the collective curvature radiation}
\author[Ya. N. Istomin, A. A. Philippov, and V. S. Beskin]
{Ya. N. Istomin$^{1,2}$, A. A. Philippov$^{2}$ and V. S. Beskin$^{1,2}$ \\
$^{1}$P.N.Lebedev Physical Institute, Leninsky prosp., 53, Moscow, 119991, Russia\\
$^{2}$Moscow Institute of Physics and Technology, Dolgoprudny,
Moscow region, 141700, Russia}

\begin{document}

\date{Accepted, Received}

\maketitle

\label{firstpage}

\begin{abstract}
The paper deals with the one possible mechanism of the pulsar radio emission, i.e., with
the collective curvature radiation of the relativistic particle stream moving along the
curved magnetospheric magnetic field lines. It is shown that the electromagnetic wave
containing one cylindrical harmonic $\exp\{is\phi\}$ can not be radiated by the curvature
radiation mechanism, that corresponds to radiation of a charged particle moving along curved
magnetic field lines. The point is that the particle in vacuum radiates the triplex of
harmonics ($s, s\pm 1$), so for the collective curvature radiation the wave polarization
is very important and cannot be fixed a priori. For this reason the polarization of real
unstable waves must be determined directly from the solution of wave equations for the media.
Its electromagnetic properties should be described by the dielectric permittivity tensor
${\hat \varepsilon}(\omega, {\bf k,r})$, that contains the information on the reaction on all possible
types of radiation.
\end{abstract}


\section{Introduction}
The curvature radiation is the type of the bremsstrahlung radiation when
a radiated charged particle moves along the curved trajectory with the
curvature radius $\rho_0$ and its acceleration is orthogonal to the velocity ${\bf v}$.
The cyclotron rotation of a charged particle in the external magnetic field $B$
is the example of this motion when $\rho_0=v_\perp/\omega_{\rm c}$. Here $\omega_{\rm c}$ is the
cyclotron frequency, $\omega_{\rm c}=eB/{m_{e}}c\gamma$, $e$ and $m_{\rm e}$ are the charge and the
mass of a particle, $\gamma$ is the particle Lorentz factor, and
$v_\perp$ is the component of the particle velocity which
is orthogonal to the magnetic field.

Moving along the circular trajectory, a particle radiates at
harmonics of the cyclotron frequency: $\omega=n\omega_{\rm c}$.
This radiation is called as cyclotron radiation for a
nonrelativistic particle and as synchrotron radiation for a
relativistic particle ($\gamma \gg 1$). For the synchrotron
radiation the maximum of the radiated power turns to be at large
numbers of cyclotron harmonics: $n \simeq \gamma^3$. The total
radiation power $I$ also grows with the particle energy,
$I\propto\gamma^2$. Therefore, the synchrotron radiation of
relativistic particles is presented widely in the space radiation
(Ginzburg \& Syrovatskii 1964).

It is necessary to stress that the length of formation of the curvature
radiation, though is larger than the wave length $\lambda$, is much less than curvature
radius $\rho_0$. So, the properties of the curvature radiation do not differ from
that of the synchrotron radiation in which the cyclotron radius is equal to
the local curvature radius $\rho_0$. The frequency of the maximum of the spectral
power, $\omega \simeq c\gamma^3/\rho_0$, and radiation power,
$I=4/3 \, e^2 c\gamma^4/\rho_0^2$, increases with the particle energy. Here the
dependence
on $\gamma$ is stronger than for the synchrotron radiation because the
curvature is fixed, and does not fall with the energy as for the motion in the
constant magnetic field.

The curvature mechanism of radiation is believed to be connected
with the mechanism of the coherent pulsar radio emission. Indeed,
in the region of the open magnetic field lines in the pulsar
magnetosphere there is a relativistic electron-positron plasma
moving with relativistic velocities along curved magnetic field
lines. For the typical values of curvature radius, $\rho_0 \simeq
10^8$ cm, and Lorentz factor of electrons and positrons,
$\gamma\simeq 10^2$, the characteristic frequency of the curvature
radiation is in the radio band. In the magnetosphere, where the
curvature frequency coincides with the plasma frequency,
$\omega_{\rm p}/\gamma^{3/2}$, we can expect
 the collective curvature radiation. Here $\omega_{\rm p}=(4\pi e^2n_{\rm e}/m_{\rm e})^{1/2}$ is the usual
plasma frequency, and $n_{\rm e}$ is the plasma number density. In the strong magnetic field
of the pulsar magnetosphere, when the charged particles can move along magnetic field
lines only, the frequency of the plasma oscillations is $\gamma^{-3/2}$ times
less than the usual plasma frequency $\omega_{\rm p}$.

It seems natural to continue the analogy between the curvature
radiation and the cyclotron radiation for the collective
radiation. But there is the essential difference between them,
which does not permit to rewrite formulas of cyclotron plasma
radiation for the curvature radiation replacing the cyclotron
radius by the curvature radius. The matter is that at each point
of plasma in the magnetic field the distribution of particles over
transverse velocities is isotropic. All directions of particle
transverse motion exist, so that the average velocity equals zero.
It is not so for the curvature radiation when all particles have
only one direction of motion along the magnetic field.

For the plasma physics the problem of the collective curvature
radiation is rather complicated since it demands the consideration
of an essentially nonuniform plasma. It does not result from the
change of parameters of magnetic field and plasma in space. These
effects can be taken into account in the local approximation
because the wave length of radiation is much less than the scales
of inhomogeneities. In order not to lose the curvature radiation
we need to include into consideration the turn of the vector of
anisotropy of the particle distribution function, $f({\bf
p})\propto\delta({\bf p}-p_\parallel {\bf B}/B)$, in space. Here
$p_\parallel$ is the longitudinal particle momentum.

Two parameters of the curvature radiation, i.e., the length of formation
$l_{\rm f}=\rho_0/\gamma\simeq\lambda\gamma^2 \gg \lambda$ and the width
of the radiation directivity $\delta\phi\simeq\gamma^{-1}$, connect by the
relation $l_{\rm f}/\rho_0\simeq\delta\phi$. Thus, the particle is in the
synchronism with the wave (i.e., the particle sees the constant wave phase)
along the path on which the wave intensity changes essentially. The value
of $\lambda$ is the wave length of the curvature radiation,
$\lambda\simeq\rho_0/\gamma^3$.

The problem of calculation of the dielectric permittivity in the
geometrical optics approximation for the nonuniform anisotropic
plasma particle distribution was solved by Beskin, Gurevich and
Istomin (below BGI, 1993). They also described the collective
curvature-plasma interaction, when the electromagnetic waves,
connected with the curvature radiation, are amplified
simultaneously by the Cherenkov mechanism. This effect is absent
in the vacuum. However, this procedure is rather complicated and
demands clear understanding. Because of that there are some
incorrect statements in the literature (see, e.g., Nambu 1989;
Machabeli 1991, 1995).

Apart, another way of investigation of the problem of the
collective curvature radiation was carried out during many years
(Asseo et al. 1983; Larroche \& Pellat 1987; Lyutikov et al. 1999;
Kaganovich \& Lyubarsky 2010). They considered more simple task
connected with the pure cylindrical geometry which can be solved
"exactly". In such a statement the magnetic field lines are
considered to be concentric, the relativistic plasma moving (i.e.,
rotating) along the magnetic field lines owing to the centrifugal
drift directed parallel to the cylindrical axis ($z$-coordinate)
with the velocity $u = c\rho_c/\rho_0 \ll c$. Here again
$\rho_c=c/\omega_{\rm c}$. But this approach cannot be used when
analyse the curvature radiation (Beskin, Gurevich \& Istomin
1988).

Indeed, let us choose the electromagnetic fields of the wave, as was done in all the papers
mentioned above, in the form
\begin{equation}
\left({\bf E},{\bf B}\right)=\left({\bf E}(\rho),{\bf B}(\rho)\right)\times\exp\left\{-i
\omega t+is\phi+ik_z z\right\}.
\end{equation}
Here $\omega$ is the wave frequency, $s$ is integer number
defining the azimuthal wave vector $k_\phi$, and $k_z$ is the
longitudinal wave vector along the cylinder. In this approach the
wave amplitudes ${\bf E}(\rho), {\bf B}(\rho)$ are to be
considered as functions of the radial distance $\rho$ only.
Moreover, not vectors ${\bf E}$ and ${\bf B}$, but their
cylindrical components $(E,B)_\rho, (E,B)_\phi$ and $(E,B)_z$
depend on the coordinate $\rho$ only. It means that the wave
polarization follows the magnetic field, turning from one point
$\phi$ to another. It can be so if we have the definite boundary
condition, e.g., putting the system into the metallic coat. Under
such suggestions we come to the one dimensional problem, which can
be easily solved. Here we will show that such a wave does not have
any relation to the curvature radiation.

Really, let us consider the particle moving exactly along the
circle of radius $\rho_0$ with the constant velocity $v$; this
motion corresponds to the infinite magnetic field. Then the
radiated power is equal to the work of the wave electric field
under the particle electric current. The electric current is
\begin{equation}
{\bf j}= e v \delta(\phi-\Omega t)\, \delta(z)
\frac{\delta(\rho-\rho_0)}{\rho}\, {\bf e}_\phi,
\end{equation}
where $\Omega=v/\rho_0$, and for selected polarization we get
\begin{equation}
\int {\bf jE}{\rm d}{\bf r}=evE_\phi(\rho_0)\exp\left\{-i\omega t+is\Omega t\right\}.
\end{equation}
As we see, the radiation is possible only if $\omega-s\Omega=0$,
i.e., $\omega=k_\phi v$. It is just the condition of Cherenkov,
not curvature radiation. The point is that the wave with such
polarization can not be radiated by the curvature mechanism. The
difference between the curvature wave and the Cherenkov wave is in
the finite interaction time of the bremsstrahlung radiation with a
radiated particle. The freely propagating wave with almost
constant polarization deflects from the direction of a particle
motion. As a result, the nonzero projection of the wave electric
field on the particle velocity (i.e., on the direction of the
electric current) occurs, and the wave takes away the energy from
the particle. This continues the finite time $\tau=l_f/v$ that can
be determined from the relation $\tau(\omega-{\bf kv})\simeq 1$.
For the relativistic particle ($v\simeq c$) $\tau=(\rho_0^2/\omega
c^2)^{1/3}\simeq \rho_0/c\gamma$. Below we will find the real
polarization of the curvature radiation.

The paper is organized as follows. In section 2 we will find that the
polarization of the curvature wave does not correspond to one cylindrical
harmonic. In section 3 it is shown that the nonlinear wave interaction
can lead to significant changes in cylindrical modes propagation.
In section 4 the BGI permittivity tensor will be derived from the permittivity
corresponding to one cylindrical mode. Finally, in section 5 we discuss the
main results of our consideration.

\section{Polarization of the curvature wave}

The radiation field of the electric current density ${\bf j}$ and the electric
charge density $\rho_{\rm e}$ of the moving particle with the charge $e$
is described by the retarded potentials (Landau \& Lifshits 1975):
\begin{eqnarray}
{\bf A} & = & \frac{1}{c}\int\frac{{\bf j}(t')}{R}{\rm d}{\bf r},
\label{A}\\
\Phi & = & \int\frac{\rho_{\rm e}(t')}{R}{\rm d}{\bf r}.
\label{Q}
\end{eqnarray}
Here $t'=t-R/c$ is the retarded time, $R$ is the distance from the charge location
at the time $t'$ to the observer which has the cylindrical coordinates $(\rho, \phi, z)$,
\begin{eqnarray}
R & = & \left[\rho^2+z^2+\rho_0^2-2\rho\rho_0\cos(\phi'-\phi)\right]^{1/2}, \\
\phi' & = & \Omega t'.
\end{eqnarray}
After the Fourier transformation of potentials (\ref{A})--(\ref{Q}) over the time
we obtain
\begin{eqnarray}
{\bf A}_\omega & = & \frac{1}{2\pi}\int{\bf A}(t)\exp\{i\omega t\}{\rm d}t, \\
\Phi_\omega & = & \frac{1}{2\pi}\int\Phi(t)\exp\{i\omega t\}{\rm d}t.
\end{eqnarray}
It is convenient now to replace the integration over time $t$ by the integration over
the retarded time $t'$ and then over the angle $(\phi'-\phi)$. As a result one can obtain
for the  cartesian components ($x,y,z$) of the vector potential ${\bf A}$ and the scalar
potential $\Phi$
\begin{equation}
\left[{\bf A}_{\omega};\,\Phi_{\omega}\right]=\frac{e\rho_0}{2\pi c}\exp\{i\omega\phi/\Omega\}
\left[-K_s;\,K_c;\, 0;\,\frac{c}{v}K_0\right].
\label{eqn1}
\end{equation}
Here the quantities $K_0, K_s, K_c$ are the functions of coordinates $\rho$ and $z$ only and
they are equal to
\begin{eqnarray}
K_0 & =& \int\frac{\exp\{i\omega(R/c+\Omega^{-1})\alpha\}}
{R+v\rho\sin\alpha/c}{\rm d}\alpha, \nonumber \\
K_s & = & \int\frac{\exp\{i\omega(R/c+\Omega^{-1})\alpha\}\sin\alpha}
{R+v\rho\sin\alpha/c}{\rm d}\alpha, \\
K_c & =& \int\frac{\exp\{i\omega(R/c+\Omega^{-1})\alpha\}\cos\alpha}
{R+v\rho\sin\alpha/c}{\rm d}\alpha, \nonumber \\
R & = & (\rho^2+z^2+\rho_0^2-2\rho\rho_0\cos\alpha)^{1/2}. \nonumber
\end{eqnarray}
The expression (\ref{eqn1}) is valid at any point ${\bf r}$, i.e., not only in
the wave zone. The dependence over the angle $\phi$ is given by the
exponent $\exp\{i\omega\phi/\Omega\}$. From the periodicity over $\phi$ we
have $\omega=s\Omega$.

The key point of the above expansion (\ref{eqn1}) is that the radiated wave is the
superposition of three harmonics: $s$,$s-1$, and $s+1$. For example, the
azimuthal electric field $E_{\phi\omega}$ is equal to
\begin{eqnarray}
&&E_{\phi\omega} = \frac{i\omega}{v}\left(-\frac{\rho_0}{\rho}\Phi_\omega
+\frac{v}{c}A_{\phi\omega}\right) \nonumber \\
&&= -i\frac{e\rho_0\omega}{2\pi v^2}e^{is\phi}\left[
\frac{\rho_0}{\rho}K_0-\frac{v^2}{c^2}\left(K_s\sin\phi + K_c\cos\phi\right)
\right].
\label{eqn6}
\end{eqnarray}
The first term in Eqn. (\ref{eqn6}), which is proportional to the
scalar potential $\Phi$, is not important in the wave zone, $\rho
\gg \rho_0$, but is significant in the near zone on the particle
trajectory $\rho=\rho_0$. Due to this term, the particle, which is
in the resonance with one of three harmonics, say with $s$
($\omega= s\Omega$), is beaten out of the synchronism by neighbour
harmonics $s\pm 1$. The electric field $E_{\phi\omega}$ changes
its sign during the time $\tau$. The synchronism condition, i.e.,
$1-\cos\Omega\tau\simeq 1-v^2/c^2=\gamma^{-2}$, defines the time
$\tau$,
\begin{equation}
\tau\simeq 1/\Omega\gamma=\rho_0/c\gamma,
\end{equation}
which coincides with the time of formation of the curvature radiation.

Thus, the radiated curvature wave consists of three harmonics $s,
s\pm 1$ with the fixed relation between their amplitudes. Namely,
this circumstance provides the curvature mechanism of the
radiation. Appearance of harmonics $s\pm 1$ except the resonant
one $s=\omega/\Omega$ is due to the additional modulation of the
radiation field induced by a modulation of the particle electric
current having the harmonic $s=1$. Now one can understand why the
simple problem of the collective curvature radiation in the
cylindrical geometry with only one azimuthal harmonic $\exp\{i s
\phi\}$ does not reveal any significant amplification of waves
(Asseo et al., 1983; Lyutikov et al., 1999; Kaganovich \&
Lyubarsky, 2010). In this case the chosen wave polarization does
not contain primordially the curvature mechanism.

\section{Collective triple radiation}

In the previous section it was shown that the curvature radiation
of one charged particle can not be described in the pure
cylindrical geometry by one azimuthal harmonic $\exp\{is\phi\}$.
In a collective radiation the modulation of the particle electric
current appears together with electromagnetic field excitation.
Because of that the resonant azimuthal harmonic
$s=\omega\rho/v_\phi$ mixes with harmonics of the electric current
modulation and produces all possible values of $s$. Further in the
section 4 we will see the all azimuthal harmonics $s$ give
contribution to the response of a media on an electromagnetic
field. But in this section it will be demonstrated that the
collective curvature radiation of only triplex of azimuthal
harmonics $(s, s\pm 1)$ differs significantly from that of one
harmonic $s$ as it is usually considered in the literature.

Let us consider the simple cylindrical one-dimensional problem of
radiation of the cold stream of plasma particles with the charge
$e$ and the mass $m_{\rm e}$ moving along the infinite azimuthal
magnetic field $B_0 = B_{\phi}$. In this case the particles can
move only in $\phi$-direction with the velocity $v_{\phi}$ at
different cylindrical radius $\rho$. The unperturbed particle
density $n^{(0)}$ and velocity $v_\phi^{(0)}$ are constants, i.e.,
they do not depend on $\rho$. The electric current ${\bf j}$ has
only $\phi$-component as well as $B_z$-component of the wave
magnetic field ($B_\rho = B_\phi=0$). Accordingly, the wave
electric field has two components $E_\rho$ and $E_\phi$ ($E_z=0$).

The dependence of the wave fields over time and coordinates is the following
\begin{equation}
[E_{\rho}; \, E_{\phi}; \, B_{z}] = [E_{\rho}(\rho); \, E_{\phi}(\rho); \, B_{z}(\rho)]\exp\{-i\omega t +is\phi\}.
\end{equation}
Then, we obtain from Maxwell equations
\begin{eqnarray}
\frac{{\rm d} E^{(\sigma)}_{\phi}}{{\rm d} \rho} = \frac{i \sigma}{\rho} E^{(\sigma)}_{\rho} -
i \frac{\rho}{\sigma}\frac{\omega^2}{c^2} E^{(\sigma)}_{\rho} - \frac{E^{(\sigma)}_{\phi}}{\rho},\\
\frac{{\rm d} E^{(\sigma)}_{\rho}}{{\rm d} \rho} = - i \frac{\sigma}{\rho} E^{(\sigma)}_{\phi} +
\frac{4 \pi}{\omega}\frac{\sigma}{\rho} j^{(\sigma)}_{\phi} - \frac{E^{(\sigma)}_{\rho}}{\rho}.
\end{eqnarray}
Here index $\sigma$ corresponds to one of three harmonics $s$ or
$s\pm 1$. For simplicity we use here the dimensionless variable
$r$ defined as $r = \rho\omega/c$, as well as quantities $\Lambda
= \omega^2_{p}/(\omega^2\gamma^3)$ and $ J_{\sigma} = 4 \pi
j^{(\sigma)}_{\phi}/(\Lambda \omega)$. Here $\omega_{\rm p}=(4\pi
ne^2/m_{e})^{1/2}$ is plasma frequency,  and $\gamma$ is the
Lorentz-factor of the particle motion:
$\gamma=(1-v_\phi^2/c^2)^{-1/2}$. After these definitions the
equations above take the following form
\begin{eqnarray}
\frac{{\rm d} E^{(\sigma)}_{\phi}}{{\rm d} r} & = & \frac{i \sigma}{r} E^{(\sigma)}_{\rho} - i \frac{r}{\sigma}
E^{(\sigma)}_{\rho} - \frac{E^{(\sigma)}_{\phi}}{r},\label{eq78}\\
\frac{{\rm d} E^{(\sigma)}_{\rho}}{{\rm d} r} & = & - i \frac{\sigma}{r} E^{(\sigma)}_{\phi} +
\Lambda \frac{\sigma}{r} J_{\sigma} - \frac{E^{(\sigma)}_{\rho}}{r}.
\end{eqnarray}

As was already stressed, we consider here the interaction of three
waves
 $s,s\pm 1$. It is important that they are not independent and their
interaction  is realized by the static electric field
$[E_\rho(\rho); \, E_\phi(\rho)] \exp\{i\phi\}$ having the first
azimuthal harmonic $s=1$. This electrostatic field turns to be the
result of nonlinear interactions of high frequency neighbour
harmonics $s$ and $s\pm 1$. Equations for the mode $s = 1$ under
the same definitions are
\begin{eqnarray}
\frac{{\rm d} E_{\phi}}{{\rm d} r} & = & \frac{i}{r} E_{\rho} - \frac{E_{\phi}}{r},\\
\frac{{\rm d} E_{\rho}}{{\rm d} r} & = & - i \frac{1}{r} E_{\phi} + \Lambda Z - \frac{E_{\rho}}{r}.
\label{eq90}
\end{eqnarray}
Here $ Z = 4 \pi n e c/(\Lambda \omega)$.

To determine the response of the stream on the electromagnetic fields of the wave one can use
the continuity and Euler equations
\begin{eqnarray}
&&\frac{\partial n}{\partial t}+{\rm \nabla}(n {{\bf v}})=0,\\
&&\left(\frac{\partial}{\partial t}+{\bf v}\nabla\right){\bf p}
=e\left({\bf E}+\left[\frac{{\bf v}}{c},{\bf B}\right]\right).
\end{eqnarray}
It is easily to understand that only the $\phi$-component of Euler
equation is needed, while the radial component just provides us
the equilibrium configuration across the infinite magnetic field.
We represent the plasma number density and the plasma velocity as
the expansion over  powers of the wave amplitude
\begin{eqnarray}
v_{\phi} = v^{(0)}_{\phi} +  \delta v^{(1)}_{\phi} + \delta v^{(2)}_{\phi} + ...,\\
n = n^{(0)} +  \delta n^{(1)} + \delta n^{(2)} + ....
\end{eqnarray}

The linear response can be easy found
\begin{eqnarray}
n^{(1)} & = & n^{(0)} \frac{k v^{(1)}_{\phi}}{\omega-kv_\phi^{(0)}},
\label{denom1} \\
v^{(1)}_{\phi} & = & i \frac{e E_{\phi}}{m_{\rm e} \gamma^3 (\omega-kv_\phi^{(0)})},
\label{denom2}
\end{eqnarray}
where $k=s/\rho$. On the other hand, for the nonlinear current the nonlinear relation
between  $\delta v_{\phi}$ and $\delta p_{\phi}$ should be taken into account
\begin{equation}
\delta p_{\phi} = m_{\rm e}\gamma^3 \delta v_{\phi} -
\frac{3}{2}m_{\rm e} v^{(0)}_{\phi}\gamma^5 \frac{(\delta v_{\phi})^2}{c^2}.
\end{equation}
The result of cumbersome but straightforward calculation is

\begin{widetext}

\begin{eqnarray}
J_s = \frac{1}{1- s v^{(0)}_{\phi}/r}\left[i\frac{E^{s}_{\phi}}
{1- s v^{(0)}_{\phi}/r} + \alpha\frac{r}{v^{(0)}_{\phi}}\left(A_{s,s-1}\frac{E^{s-1}_{\phi}E^{1}_{\phi}}{1- (s -1)
v^{(0)}_{\phi}/r} -
A_{s,s+1}\frac{E^{s+1}_{\phi}E^{1*}_{\phi}}{1- (s +1) v^{(0)}_{\phi}/r}\right)\right],\label{Jeq1}\\
J_{s-1} = \frac{1}{1- (s -1) v^{(0)}_{\phi}/r}\left[i\frac{E^{s-1}_{\phi}}{1- (s -1) v^{(0)}_{\phi}/r} - \alpha\frac{r}{v^{(0)}_{\phi}}A_{s,s-1}\frac{E^{s}_{\phi}E^{1*}_{\phi}}{1- s v^{(0)}_{\phi}/r}\right],\\
J_{s+1} = \frac{1}{1- (s +1) v^{(0)}_{\phi}/r}\left[i\frac{E^{s}_{\phi}}{1- (s +1) v^{(0)}_{\phi}/r} + \alpha\frac{r}{v^{(0)}_{\phi}}A_{s,s+1}\frac{E^{s}_{\phi}E^{1}_{\phi}}{1- s v^{(0)}_{\phi}/r}\right],\\
Z = \frac{1}{\left(v^{(0)}_{\phi}\right)^2} \left[i\frac{E_{1}}{1/r} + \alpha \left(\frac{E^{s+1}_{\phi}E^{s*}_{\phi}}{(1- (s +1) v^{(0)}_{\phi}/r)(1- s v^{(0)}_{\phi}/r)} + \frac{E^{s}_{\phi}E^{(s-1)*}_{\phi}}{(1- s v^{(0)}_{\phi}/r)(1- (s -1) v^{(0)}_{\phi}/r)}\right)\right],\label{Jeq2}\\
A_{i,j} = \frac{1}{1- i v^{(0)}_{\phi}/r} + \frac{1}{1- j v^{(0)}_{\phi}/r} -3 \gamma^2,\nonumber
\end{eqnarray}
\end{widetext}
Here $\alpha = e/(m_{e}c\gamma^3\omega)$ is the particle velocity
divided over the velocity of light. The same quantities for plane
waves can be found in (BGI, 1993). Equations above are evaluated
with vacuum initial condition for the normal mode that can be
presented analytically, $E^{(\sigma)}_{\phi} = -
J^{'}_{\sigma}(r), E^{(\sigma)}_{r} = i {\sigma} J_{\sigma}(r)/r$.
Here $J_\sigma(r)$ is the Bessel function. It should be noted that
the singularity in equations (\ref{eq78}) is passed smoothly by
additional small term $ +i \varepsilon$ in the resonance
denominators in (\ref{denom1})--(\ref{denom2}).

\begin{figure*}
\includegraphics[scale=0.3]{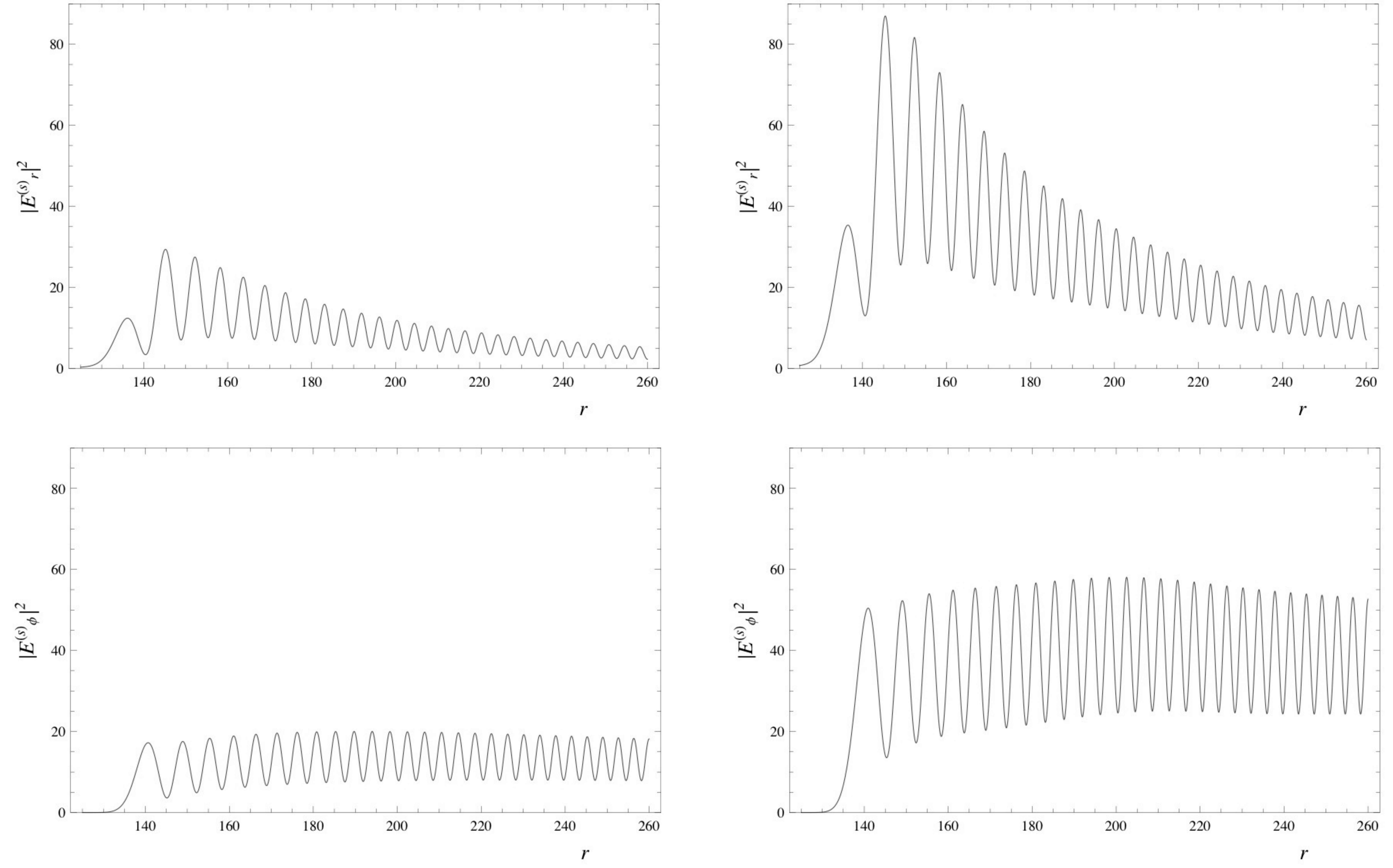}
\caption{Model calculations of two cases, $\Lambda = 10^{-2}$, $\nu = 1 GHz$, $\gamma =5$, $s = 125$.}
\label{fig1}
\end{figure*}

In numerical calculations equations (\ref{eq78})--(\ref{eq90}) for $\sigma = s$ and $\sigma = s \pm 1$
were solved with two different values for the quantities $J_{\sigma}$ and $Z$. In the first case
we neglect non-linear terms in (\ref{Jeq1})--(\ref{Jeq2}), while the second one corresponds
to the full non-linear problem. On Fig. ~\ref{fig1} the results obtained for this cases are
presented. For better representation of the influence of the nonlinear current, we choose
the amplitudes of $s - 1$ and $s + 1$ modes twenty times higher than the amplitude
of the $s$ mode. In reality the $s$-mode interacts with the whole continuum of modes,
so this model assumption is rather reasonable. Fig. ~\ref{fig1} shows that in this case
the intensity of the wave $|E|^2$ is approximate 2.5 times larger than in the case when
the nonlinear current is neglected. Hence, one can conclude that three wave interaction
is rather effective.

Thus, we have shown that the triplex of cylindrical harmonics,
which corresponds better to the curvature mechanism, is amplified
more effective than the separated harmonic having the single value
of the azimuthal number. In fact the real polarization of the
collective curvature mode can be obtained only by calculating the
permittivity tensor of the streaming plasma in the strong curved
magnetic field. The solution of wave equations produces not only
the dispersive equation for normal waves, $\omega =\omega({\bf
k})$, but defines also their polarization. A priory it is unclear
what polarization corresponds to unstable modes.

At first sight, the problem considered above is essentially
nonlinear and has no direct connection with the question of the
linear wave amplification. We included nonlinearity only in order
to connect harmonics $s,\, s\pm 1$ self-consistently. And
appearance of neighbour harmonics $s\pm 1$, even for small
nonlinearity, strongly change the $s$-mode amplification. It is
clear also that interaction of $s\pm 1$ waves with the field $s=1$
will result in all azimuthal harmonics.

\section{Tensor derivation}

In this section we will show that the asymptotic behaviour of the BGI dielectric tensor in the
case of large enough curvature radius $\rho_{0}$ can be found directly from the plasma response on
the one cylindrical mode. For the infinite toroidal magnetic field
only the response to the toroidal component of the wave electric field $E_{\phi}$ is to be
included into consideration (Beskin 1999). Here and below we consider the stationary medium
only, so the time dependence can be chosen as $\exp\{- i \omega t\}$. Making summation over
all cylindrical modes, one can write down
\begin{equation}
D_{\phi} (\rho, \phi) = E_{\phi} (\rho, \phi) -
\sum\limits_{s = - \infty}^{\infty} E_{\phi}(\rho, s) K(\rho, s) \exp\{is\phi\},
\label{eqm}
\end{equation}
where
\begin{equation}
K (\rho, s) = \frac{4 \pi e^2}{\omega} \int
\frac{v_{\phi}}{\omega - s v_{\phi}/\rho} \frac{\partial f^{(0)}}{\partial p_{\phi}}
{\rm d}p_{\phi}.
\end{equation}
Here $f^{(0)}(p_{\phi})$ is unperturbed distribution function.
Making the Fourier transformation
\begin{equation}
E_{\phi}(\rho,s) = \frac{1}{2\pi} \int\limits_{0}^{2\pi} E_{\phi}(\rho,\phi') \exp\{- i s\phi'\} {\rm d}\phi'
\end{equation}
and the transition to cartesian coordinate system one can obtain:
\begin{eqnarray}
D_{x} = E_{x} + \frac{1}{2\pi}\int\frac{\rho'{\rm d}\rho'{\rm d}\phi'}{\rho'}
\sum\limits_{s = - \infty}^{\infty}E_{\phi}(\rho', \phi')\delta(\rho - \rho')
\times\nonumber \\
K(\rho, s) \, \exp\{i s (\phi - \phi')\}\sin \phi,\\
D_{y} = E_{y} - \frac{1}{2\pi}\int\frac{\rho'{\rm d}\rho'{\rm d}\phi'}{\rho'}
\sum\limits_{s = - \infty}^{\infty}E_{\phi}(\rho', \phi')\delta(\rho - \rho')
\times \nonumber \\
K(\rho, s) \, \exp\{i s (\phi - \phi')\}\cos \phi.
\end{eqnarray}
We choose the local coordinate system with the $y$-axis directed
along the magnetic field and the $x$-axis, that is orthogonal to
it. From the above equations one can obtain the permittivity
kernel components
\begin{eqnarray}
\varepsilon_{yy}({\bf{r}}, {\bf{r'}}) = 1 - \frac{1}{2\pi}\frac{1}{\rho'}
\sum\limits_{s = - \infty}^{\infty}\delta(\rho - \rho') K(\rho, s)\times
\nonumber  \\
\exp\{i s (\phi - \phi')\}\cos \phi \cos \phi';\\
\varepsilon_{yx}({\bf{r}}, {\bf{r'}}) = \frac{1}{2\pi}\frac{1}{\rho'}
\sum\limits_{s = - \infty}^{\infty}\delta(\rho - \rho') K(\rho, s) \times
\nonumber  \\
\exp\{i s (\phi - \phi')\}\cos \phi \sin \phi';\\
\varepsilon_{xy}({\bf{r}}, {\bf{r'}}) = \frac{1}{2\pi}\frac{1}{\rho'}
\sum\limits_{s = - \infty}^{\infty}\delta(\rho - \rho') K(\rho, s)\times
\nonumber \\
\exp\{i s (\phi - \phi')\}\sin \phi \cos \phi',\\
\varepsilon_{xx}({\bf{r}}, {\bf{r'}}) = 1 - \frac{1}{2\pi}\frac{1}{\rho'}
\sum\limits_{s = - \infty}^{\infty}\delta(\rho - \rho') K(\rho, s)\times\nonumber \\
\exp\{i s (\phi - \phi')\}\sin \phi \sin \phi',
\end{eqnarray}
that provides the material relationship
\begin{equation}
D_{i}({\bf{r}}) = \int \varepsilon_{ij}({\bf{r}}, {\bf{r'}})E_{j}({\bf{r'}}) {\rm d} {\bf{r'}}.
\end{equation}

It should be noted that the operator presented above satisfies the needed symmetry condition
\begin{equation}
\varepsilon_{ij}({\bf{r}}, {\bf{r'}}, \omega) = \varepsilon_{ji} ({\bf{r'}}, {\bf{r}}, -\omega)
\end{equation}
(it is provided by the condition $K(r, s, \omega) = K(r, -s, -\omega)$). As it is well-known
(Kadomtsev 1965; Bornatici \& Kravtsov 2000), it is this symmetrical form of permittivity tensor
that is to be used for the calculation of components of the permittivity tensor
$\varepsilon_{ij} (\omega, {\bf {k}}, {\bf {r}})$
\begin{equation}
\varepsilon_{ij} (\omega, {\bf {k}}, {\bf {\boldsymbol\eta}} \to {\bf {r}})
= \int \varepsilon_{ij}(\omega, {\bf{\boldsymbol\xi}}, {\bf{\boldsymbol\eta}}) \exp\{-i\bf{k\boldsymbol\xi}\}{\rm d} {\bf{r}} .
\label{kadom}
\end{equation}
Here  ${\boldsymbol{ \boldsymbol\eta}} = ({\bf r} + {\bf r'})/2$,
${\bf \boldsymbol\xi} = \bf{r} - \bf{r'}$. It is important that
the above tensor only describes correctly wave-particle
interaction in inhomogeneous media with slowly varying parameters
(Bernstein \& Friedland 1984).

Substituting now the kernel components, one can find
\begin{eqnarray}
\varepsilon_{xx}(\omega, {\bf {k}}, {\bf {\boldsymbol\eta}}) = 1 - \frac{1}{2\pi}\int {\rm d}{\bf \boldsymbol\xi} \exp\{-i{\bf{k \boldsymbol\xi}\}}\frac{1}{|{\bf \boldsymbol\eta - \boldsymbol\xi}/2|}\times\nonumber\\\sum\limits_{s = - \infty}^{\infty}\delta(|{\bf \boldsymbol\eta + \boldsymbol\xi}/2| -|{\bf \boldsymbol\eta - \boldsymbol\xi}/2|) K(|{\bf \boldsymbol\eta + \boldsymbol\xi}/2|, s) \times
\nonumber\\
\exp\{i s (\phi - \phi')\}\sin \phi \sin \phi',
\label{eps1} \\
\varepsilon_{xy}(\omega, {\bf {k}}, {\bf {\boldsymbol\eta}}) = \frac{1}{2\pi}\int {\rm d}{\bf \boldsymbol\xi} \exp\{-i{\bf{k \boldsymbol\xi}\}}\frac{1}{|{\bf \boldsymbol\eta - \boldsymbol\xi}/2|}\times\nonumber\\\sum\limits_{s = - \infty}^{\infty}\delta(|{\bf \boldsymbol\eta + \boldsymbol\xi}/2| -|{\bf \boldsymbol\eta - \boldsymbol\xi}/2|) K(|{\bf \boldsymbol\eta + \boldsymbol\xi}/2|, s)\times\nonumber \\
\exp\{i s (\phi - \phi')\}\sin \phi \cos \phi',
\label{eps12} \\
\varepsilon_{yx}(\omega, {\bf {k}}, {\bf {\boldsymbol\eta}}) = \frac{1}{2\pi}\int {\rm d}{\bf \boldsymbol\xi} \exp\{-i{\bf{k \boldsymbol\xi}\}}\frac{1}{|{\bf \boldsymbol\eta - \boldsymbol\xi}/2|}\times\nonumber\\\sum\limits_{s = - \infty}^{\infty}\delta(|{\bf \boldsymbol\eta + \boldsymbol\xi}/2| -|{\bf \boldsymbol\eta - \boldsymbol\xi}/2|) K(|{\bf \boldsymbol\eta + \boldsymbol\xi}/2|, s)\times\nonumber \\
\exp\{i s (\phi - \phi')\}\cos \phi \sin \phi',
\label{eps21} \\
\varepsilon_{yy}(\omega, {\bf {k}}, {\bf {\boldsymbol\eta}}) = 1 -\frac{1}{2\pi}\int {\rm d}{\bf \boldsymbol\xi} \exp\{-i{\bf{k \boldsymbol\xi}\}}\frac{1}{|{\bf \boldsymbol\eta - \boldsymbol\xi}/2|}\times\nonumber\\\sum\limits_{s = - \infty}^{\infty}\delta(|{\bf \boldsymbol\eta + \boldsymbol\xi}/2| -|{\bf \boldsymbol\eta - \boldsymbol\xi}/2|) K(|{\bf \boldsymbol\eta + \boldsymbol\xi}/2|, s) \times\nonumber \\
\exp\{i s (\phi - \phi')\}\cos \phi \cos \phi'.
\label{eps3}
\end{eqnarray}
In this equations, the angles $\phi$ and $\phi'$ are the functions of polar angles of vectors
$\bf{\boldsymbol\eta}$ and $\bf{\boldsymbol\xi}$, $\alpha_{\eta}$, and $\alpha_{\xi}$
\begin{eqnarray}
\sin \phi = \frac{|\boldsymbol\eta|\sin\alpha_{\eta} + (|\boldsymbol\xi|/2)\sin\alpha_{\xi}}{|{\bf \boldsymbol\eta + \boldsymbol\xi}/2|},\\
\cos \phi' = \frac{|\boldsymbol\eta|\cos\alpha_{\eta} - (|\boldsymbol\xi|/2)\cos\alpha_{\xi}}{|{\bf \boldsymbol\eta - \boldsymbol\xi}/2|}.
\end{eqnarray}
As a result, integrals above are reduced to integration over $\bf{\boldsymbol\xi}$, which is perpendicular
to $\bf{\boldsymbol\eta}$. On the other hand, the expression for the delta-functions in (\ref{eps1})--(\ref{eps3})
is the following:
\begin{eqnarray}
\delta(...) & = & \frac{\delta(\theta - \pi/2)}{(|{\bf \boldsymbol\eta + \boldsymbol\xi}/2| -|{\bf \boldsymbol\eta - \boldsymbol\xi}/2|)'_{\theta}}
\nonumber \\
& + & \frac{\delta(\theta + \pi/2)}{(|{\bf \boldsymbol\eta + \boldsymbol\xi}/2| -|{\bf \boldsymbol\eta - \boldsymbol\xi}/2|)'_{\theta}},
\label{deltaf}
\end{eqnarray}
where $\theta$ is the angle between vectors $\bf{\boldsymbol\eta}$ and ${\bf{\boldsymbol\xi}}$. So, the integration
over angles can be done easily. Finally, from the transition ${\bf {\boldsymbol\eta}} \to {\bf {r}}$, one
can obtain $\cos\alpha_{\eta} \to \cos\alpha_{r} =1$. Hence, according to (\ref{deltaf})
$({\bf k \boldsymbol\xi}) = k_{\parallel} |\boldsymbol\xi|$, where $k_{\parallel}$ is the component of the wave vector
parallel to external magnetic field.

The property of the absence of $k_{\perp}$ is very important, it provides the same symmetry as
it was in the case of homogeneous medium: $\varepsilon_{ij}(- \omega, -{\bf k}, -{\bf B}, {\bf r}) =
\varepsilon_{ji}(\omega, {\bf k}, {\bf B}, {\bf r})$ (Istomin 1994). This result differs from
one obtained by Lyutikov at al. (1999). In this work the importance of transformation (\ref{kadom})
is neglected.

Using finally the Taylor expansion over $|\boldsymbol\xi|$ and the reduction of resonant denominator to delta-function,
one can obtain:
\begin{eqnarray}
&& \sum(...)\frac{1}{\omega|{\bf \boldsymbol\eta + \boldsymbol\xi}/2|/v_{\phi} - s}
\nonumber \\
&& \to i\pi \int (...) \delta \left[s - \frac{\omega(|\boldsymbol\eta|^2 + |\boldsymbol\xi|^2/4)^{1/2}}{v_{\phi}}\right]{\rm d}s.
\end{eqnarray}
As a result, one can write down
\begin{eqnarray}
\varepsilon_{xx} & = & -i\frac{8\pi^2 e^2}{\omega}\int F''(\kappa) \frac{v_{\phi}}{\omega}\frac{\partial f^{(0)}}{\partial p_{\phi}}{\rm d} p_{\phi},\\
\varepsilon_{xy} & = & - \varepsilon_{yx}   =   \frac{8\pi^2 e^2}{\omega}\int F'(\kappa) \frac{\rho^{1/3}_0 v^{2/3}_{\phi}}{\omega^{2/3}}\frac{\partial f^{(0)}}{\partial p_{\phi}}{\rm d} p_{\phi},\\
\varepsilon_{yy} & = & -i\frac{8\pi^2 e^2}{\omega}\int F(\kappa) \frac{\rho^{2/3}_0 v^{ 1/3}_{\phi}}{\omega^{1/3}}\frac{\partial f^{(0)}}{\partial p_{\phi}}{\rm d} p_{\phi}.
\end{eqnarray}
Here
\begin{eqnarray}
F(\kappa) & = & \frac{1}{\pi}\int\limits^{+\infty}_{0} \exp\{i\kappa t + it^3/3\} {\rm d} t,\\
\kappa & = & \frac{2(\omega - k_{\parallel} v_{\phi})}{\omega^{1/3} v^{2/3}_{\phi}}\rho^{2/3}_{0},
\end{eqnarray}
prime means the derivative, and $\rho_0$ is the curvature radius of magnetic field.

Due to high enough curvature radius of field lines in the pulsar
magnetosphere, one can use the asymptotic behaviour of $F(\kappa)$
for $\kappa \gg 1$
\begin{equation}
F(\kappa) \approx \frac{i}{\pi \kappa} + \frac{2i}{\pi \kappa^4} + ...
\end{equation}
After integration by parts, the final result is
\begin{equation}
\varepsilon_{ij} =
\begin{pmatrix}
1 - \frac{3}{2}\left<\frac{\omega^2_{pl} v^2_{\parallel}}{\gamma^3 \rho^2_0 \tilde{\omega}^4}\right> && - i\left<\frac{\omega^2_{pl} v_{\parallel}}{\gamma^3 \rho_0 \tilde{\omega}^3}\right> \cr
 i\left<\frac{\omega^2_{pl} v_{\parallel}}{\gamma^3 \rho_0 \tilde{\omega}^3}
 \right>&& 1 -  \left<\frac{\omega^2_{pl}}{\gamma^3 \tilde{\omega}^2}\right>
 \cr
\end{pmatrix}
\label{BGItensor}
\end{equation}
Here by definition
$\tilde{\omega} = \omega - {\bf k v}$,
and the brackets $<>$ denote both the averaging over the particle distribution
function $f_{e^{+}, e^{-}}(p_{\phi})$ and the summation over the types of particles:
\begin{equation}
<(...)> \, = \sum_{e^{+}e^{-}} \int (...)  f^{(0)}_{e^{+},e^{-}}(p_{\phi}){\rm d} p_{\phi}.
\end{equation}

We see that the tensor above is just the BGI tensor, that leads to
instability of the so-called curvature plasma modes. In the limit
$\rho_{0} = \infty$ this tensor, as expected, tends to the
dielectric permittivity of a homogeneous plasma. The nonzero
components $\varepsilon_{xy}, \varepsilon_{yx}$ and
$\delta\varepsilon_{xx}=\varepsilon_{xx}-1$ in the tensor
$\varepsilon_{ij}$ (\ref{BGItensor}) for the finite curvature are
due to nonlocal properties of the plasma response on the
electromagnetic wave in curved magnetic field. The parameter of
nonlocality $(v_\parallel/\tilde{\omega})/\rho_0$ is the ratio of
the formation length of radiation to the curvature radius. For
vacuum $\tilde{\omega}\simeq\omega/\gamma^2$, and the length
$v_\parallel/\tilde{\omega}$ coincides with the length of
formation of the curvature radiation $l_f$.

It is important that the components
$\varepsilon_{xy}=-\varepsilon_{yx}$ and $\delta\varepsilon_{xx}$
essentially change the wave polarization. The relation between
$E_\phi$ and $E_\rho$ of the wave electric field, following from
the tensor of the dielectric permittivity (\ref{BGItensor}), is
\begin{equation}
\left(\varepsilon_{xy}+n_\rho n_\phi\right)E_\phi+\left(\delta\varepsilon_{xx}
+1-n_\phi^2\right)E_\rho=0,
\end{equation}
where $n_\rho$ and $n_\phi$ are components of the dimensionless wave vector:
${\bf n} = {\bf k}c/\omega$. For the tangent wave propagation (i.e., for $n_\rho=0$) we have
$E_\phi \simeq (\delta\varepsilon_{xx}/\varepsilon_{xy})E_\rho\simeq(c/\rho_c\tilde{\omega})E_\rho$.
As a result, a wave can produce the negative work under the electric particle
current $j_\phi$, i.e.,  it can be excited. It is not so if $\delta\varepsilon_{xx}=
\varepsilon_{xy}=0$ when $E_\phi=0$.

\section{Discussion}

Thus, as it was shown above, the wave polarization $[E_\rho(\rho);
\, E_\phi(\rho)]\exp\{is\phi\}$ containing one cylindrical
harmonic $s$ suggests only the Cherenkov mechanism of radiation.
In the curvature radiation mechanism of one particle in vacuum the
generated wave consists of three harmonics $s,s\pm 1$. This
property provides the exit from the phase synchronism of the wave
with the particle motion which is inherent in the bremsstrahlung
radiation. For the collective curvature radiation it is shown that
the hydrodynamical model of plasma motion along the infinite
magnetic field gives different results of the wave amplification
depending on the wave polarization. So, there is no another way to
find the polarization of exiting waves instead of calculation of
the response of the medium on an electromagnetic field, i.e., to
use the dielectric permittivity. The correct procedure of
dielectric permittivity calculation using the expansion over
cylindrical modes is also shown above. It was demonstrated that
the tensor obtained by this procedure coincides with the BGI
tensor calculated previously by another method.

In conclusion it is worth to note that unsuccessful attempts to
find the collective curvature radiation bring to the term
'curvature-drift instability' (Lyutikov et al. 1999). As was shown
the chosen simple wave polarization, i.e., one $s$-harmonic, means
only the existence of the Cherenkov mechanism of the wave
generation. In this case the centrifugal particle drift places the
significant role. Practically all curvature effects come only to
this drift. And the Cherenkov resonance on the drift motion
produces small wave amplification in better case (Kaganovich \&
Lyubarsky 2010). Stronger magnetic field produces less drift
velocity and less Cherenkov effect though the curvature of a
particle motion does not depend on the magnetic field strength at
all.

\section{Acknowledgments}

We thank A.V.~Gurevich for his interest and support.
This work was partially supported by Russian Foundation for Basic Research (Grant no.
11-02-01021).

\end{document}